\documentclass{elsart}

\usepackage{amssymb}
\usepackage{graphicx}
\usepackage{bm}

\def\vec#1{\mathchoice{\mbox{\boldmath$\displaystyle#1$}}
{\mbox{\boldmath$\textstyle#1$}}
{\mbox{\boldmath$\scriptstyle#1$}}
{\mbox{\boldmath$\scriptscriptstyle#1$}}}
\makeatletter
\newcommand\erfc{\mathop{\operator@font erfc}\nolimits}
\def\slashchar#1{\setbox0=\hbox{$#1$}
   \dimen0=\wd0 \setbox1=\hbox{/} \dimen1=\wd1
   \ifdim\dimen0>\dimen1 \rlap{\hbox to \dimen0{\hfil/\hfil}} #1
   \else  \rlap{\hbox to \dimen1{\hfil$#1$\hfil}} / \fi}

\def\w{\omega}
\makeatother

\begin{document}
\begin{frontmatter}

\title{Pion-photon Transition Distribution Amplitudes in the Spectral
Quark Model\thanksref{grant}} \thanks[grant]{Research supported in
part by the Polish Ministry of Science and Education grant
2~P03B~02828, by DGI and FEDER funds under contract FFIS2005-00810, by
the Junta de Andaluc\'\i a grant FM-225, and EURIDICE grant
HPRN-CT-2003-00311.}

\author[as,ifj]{Wojciech Broniowski},
\ead{Wojciech.Broniowski@ifj.edu.pl}
\author[ugr]{Enrique Ruiz Arriola}
\ead{earriola@ugr.es}
\address[as]{Institute of Physics, \'Swi\c{e}tokrzyska Academy,
ul.~\'Swi\c{e}tokrzyska 15, PL-25406~Kielce, Poland} 
\address[ifj]{The H. Niewodnicza\'nski Institute of Nuclear Physics, 
Polish Academy of Sciences, PL-31342 Krak\'ow, Poland}
\address[ugr]{Departamento de F\'{\i}sica At\'omica, Molecular y
Nuclear, Universidad de Granada, E-18071 Granada, Spain} 

\begin{abstract}
The vector and axial pion-photon transition distribution amplitudes
are analyzed in the Spectral Quark Model.  We proceed by the
evaluation of double distributions through the use of a manifestly
covariant calculation based on the $\alpha$ representation of
propagators.  As a result polynomiality is incorporated and
calculations become rather simple.  Explicit formulas, holding at the
low-energy quark-model scale, are obtained.  The corresponding form
factors for the anomalous decay $\pi^0 \to \gamma \gamma^\ast$ and the
radiative pion decays $\pi^\pm \to l^\pm \nu_l \gamma$ are also
evaluated and confronted with the data.
\end{abstract}


\end{frontmatter}

\maketitle

Recently Pire and Szymanowski \cite{Pire:2004ie,Pire:2005ax} proposed
generalizations of the parton distributions to the case where the
initial and final states correspond to different particles.  Such
objects are termed {\em transition distribution amplitudes} (TDAs) and
are relevant in the analysis of the virtual Compton scattering and
other exclusive processes.  For a recent review of a related topic of
generalized parton distributions (GPDs) see, {\em e.g.},
\cite{Ji:1998pc,Golec-Biernat:1998ja,Goeke:2001tz,Diehl:2003ny,Belitsky:2005qn}
and references therein.  Of particular importance are the pion-photon
vector and axial leading-twist TDAs, $V$ and $A$, which are the
subject of this letter. A quark-model analysis of these objects has
been undertaken by Tiburzi \cite{Tiburzi:2005nj}, where the relevant
double distributions have been computed. Here we follow similar steps,
however, instead of using parameterizations \cite{Lansberg:2006fv}, we
carry out an explicit analytical calculation all the way down using
the {\em Spectral Quark Model} (SQM)~\cite{RuizArriola:2003bs}. This model
possesses a regularization which allows for a uniform treatment of both
anomalous and non-anomalous processes, essential in the study. It also
encodes the vector-meson dominance and as a result leads to very
successful phenomenology of numerous processes with pions, photons, or
$\rho$-mesons. Below we obtain simple analytic expressions for $V$ and
$A$ and the corresponding form factors, which hold at the low {\em
quark-model energy scale}.  Variants of chiral quark models have been
used to obtain information on the pion structure function
\cite{Davidson:1994uv,Frederico:1994dx}, pion distribution amplitude
\cite{Petrov:1998kg,Anikin:2000bn,Praszalowicz:2001wy,Dorokhov:2002iu,RuizArriola:2002bp},
generalized parton distributions of the pion
\cite{Polyakov:1999gs,Theussl:2002xp,Tiburzi:2002kr,Broniowski:2003rp,Noguera:2005cc},
and the photon and $\rho$ distribution amplitudes
\cite{Dorokhov:2006qm}.

In this paper the  considered TDAs are defined as \cite{Pire:2004ie}
\begin{eqnarray}
\hspace{-1cm}&& \int \! \frac{d z^-}{2\pi}\, e^{ix p^+ z^-} \!\langle
\gamma(p',\varepsilon)| \bar{\psi}(-\frac{z}{2})\gamma^\mu
\frac{\tau^a}{2}{\psi}(\frac{z}{2}) |\pi^b(p) \rangle
\Big|_{\stackrel{z^+=0}{z_T=0}} \! \nonumber \\ && =\!\frac{i e}{p^+ f} \epsilon^{\mu
\nu \alpha \beta} \varepsilon_\nu^* \, p_\alpha q_\beta V^{ab}(x,\zeta,t),
\label{VTDA} \\
\hspace{-1cm}&& \int \! \frac{d z^-}{2\pi}\, e^{ix p^+z^-} \!\langle
\gamma(p',\varepsilon)| \bar{\psi}(-\frac{z}{2})\gamma^\mu
\gamma_{5}\frac{\tau^a}{2}{\psi}(\frac{z}{2})|\pi^b(p) \rangle
\Big|_{\stackrel{z^+=0}{z_T=0}} \! \nonumber \\ &=&\! \frac{e}{p^+ f} (\varepsilon^* 
\cdot q) p'^\mu A^{ab}(x,\zeta,t)+\dots, \label{ATDA} \\
\hspace{-1cm}&& V^{ab}=\delta^{ab}V_{I=0} + i \epsilon^{ab3}V_{I=1},\;\; A^{ab}=\delta^{ab}A_{I=0} + i \epsilon^{ab3}A_{I=1},\label{isodec} 
\end{eqnarray}
where $\psi$ represents the iso-doublet quark field, $z$ is a
light-cone coordinate\footnote{We use the convention $z^{\pm}= z^0 \pm z^3 $ .} ($z^+=0$, $z_T=0$), 
$p$ denotes the momentum of the pion, $f$
is the pion decay constant with $f=86$~MeV in the chiral limit, and the
photon carries momentum $p'=p+q$ and has polarization
$\varepsilon$. Finally, in the asymmetric notation of GPDs we use
$\zeta=q^+/p^+$ and $t=q^2$. We consider {\em isovector} quark
bilinears. The isospin decomposition (\ref{isodec}) follows from the
fact that the photon couples to the quark through a combination of
isoscalar and isovector coupling, {\em i.e.} the quark charge is
$Q=1/(2N_c)+\tau^3/2$. The presence of the gauge link operators
$[-z/2,z/2]$ is understood in Eq.~(\ref{VTDA},\ref{ATDA}) in order to guarantee
gauge invariance of bilocal operators.  For brevity, in
Eq.~(\ref{ATDA}) only the piece proportional to $(\varepsilon \cdot q) p^\mu$ is
retained. The part proportional to $(\varepsilon \cdot q) q^\mu$, indicated by ellipses,
corresponds to the pion pole term in the $t$-channel and is not
relevant for the evaluation of the axial TDA (see \cite{Tiburzi:2005nj} for a
detailed discussion of the tensor structure including the pion pole
term).

\begin{figure}[tb]
\begin{center}
\includegraphics[width=11.5cm]{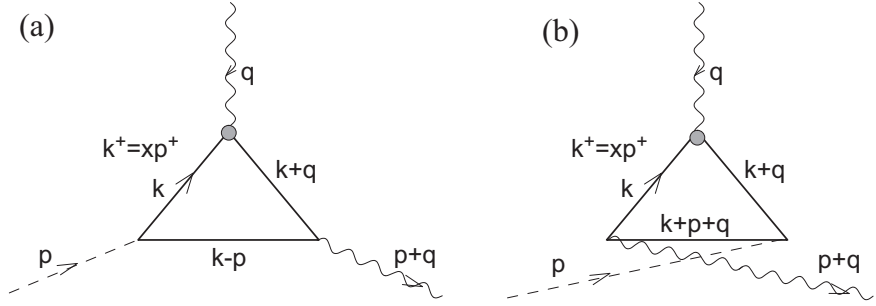} 
\end{center}
\caption{The direct (a) and crossed (b) Feynman diagrams for the quark-model
evaluation of the pion-photon TDAs. 
\label{fig:diag}}
\end{figure}

The quark-model evaluation of Eq.~(\ref{VTDA},\ref{ATDA}) amounts to
the calculation of the diagrams of Fig.~\ref{fig:diag}, where the $+$
component of the momentum of the hit quark is constrained to the value
$k^+=xP^+$. We denote $\omega$ as the constituent quark mass and use $
\gamma_5\omega/f$ as the pion-quark coupling vertex as required by the
Goldberger-Treiman relation at the quark level. We switch to the
Euclidean momenta from now on\footnote{This is not a limitation.
The calculation can be entirely carried out in the Minkowski space if the
standard boundary condition $\w^2 \to \w^2- i0^+$ is
incorporated.}. The $I=0$ and $I=1$ components are the following
combinations of the direct and crossed diagrams of
Fig.~\ref{fig:diag}: $V_{I=0}=V+\bar{V}$, $V_{I=1}=N_c(V-\bar{V})$,
where the bar indicates the crossed diagram, and similarly for
$A$. Since the crossed contributions are related to the direct terms
via a simple kinematic transformation, we first analyze the direct
diagrams.  For the vector TDA a simple algebra, involving the
evaluation of the trace factor, yields
\begin{eqnarray}
&& \hspace{-7mm} V(x,\zeta,t)= 2 \int\frac{d^4 k}{(2\pi)^4} 
\frac{\delta(k \cdot n-x)\, \omega^2}{[{k^2+\omega^2}]
[{(k+q)^2+\omega^2}] [{(k-p)^2+\omega^2}]}, \label{Vgen} 
\end{eqnarray}
where $n^2=0$, $p \cdot n=1$, and $q \cdot n=-\zeta$.  Our basic
technique makes use of the $\alpha$ representation for the scalar
propagators, which allows for a covariant treatment of the $\delta$
function. Thus,
\begin{eqnarray}
V(x,\zeta,t) &=&
2 \, \omega^2 \int\frac{d^4 k}{(2\pi)^4} \int \frac{d\lambda}{2\pi} e^{i \lambda (k \cdot n-x)} 
\label{Valpha} \\
&\times& \int_0^\infty d \alpha \int_0^\infty d \beta \int_0^\infty
d \gamma \, e^{-\alpha (k^2+\omega^2) -\beta ((k+q)^2+\omega^2) - \gamma ((k-p)^2+\omega^2) }. \nonumber
\end{eqnarray}
Next, we introduce the
variables $s=\alpha+\beta+\gamma$, $y={\beta}/{s}$, and
$z={\gamma}/{s}$, shift the momentum to $k=k'+z p - y q +i \lambda n /(2s)$, 
and then perform the Gaussian integral over $k$. Note that since $\alpha, \beta, \gamma \ge 0$, we
get $0 \le y, z \le 1$ and also $y + z \le 1$.  As a result,
\begin{eqnarray}
&& \hspace{-13mm} V\! = \!\frac{\omega^2}{8\pi^2}\! \int_0^1 \!\! d y
\!  \int_0^1 \!\! d z \theta(1\!-y\!-z) \!\int_0^\infty \!d s \delta
\left ( x\! -\! z\! -y \zeta \right ) e^{-s [\omega^2+ y(1-y)q^2
+z(1-z)p^2 + 2 y z p \cdot q]} \nonumber \\ &=& \frac{1}{8\pi^2}
\int_0^1 d y \!\int_0^1 d z \delta \left ( x\! -\! z\!-\! y \zeta
\right ) D(z,y), \label{bub}
\end{eqnarray}
where
\begin{eqnarray}
D(z,y)&=&\frac{\theta(1-y-z)\, \omega^2} {\omega^2+ y(1-y)q^2
+z(1-z)p^2 + 2 y z p \cdot q}, \label{Vdouble}
\end{eqnarray}
is the {\em double distribution} for the vector TDA. With our Euclidean vectors 
$p^2=-m_\pi^2$, $q^2=-t$, and $2 p \cdot q=m_\pi^2+t$ for the real photon. 
Equation (\ref{Vdouble}) agrees with the result of Ref.~\cite{Tiburzi:2005nj}.
Note that our use of the $\alpha$-representation has led to a direct
evaluation of TDAs from the Feynman diagrams. This is a very simple alternative 
to the more involved expansion in moments, typically done in similar
studies.

For TDAs the variable $\zeta$ cannot be assumed to be positive, as the
crossing symmetry does not relate initial and final states, unlike for
GPDs. In general we have $-1 \le \zeta \le 1$.  Next, we perform the
$z$ integration, which sets $z=x-y \zeta$. For the case $\zeta \ge 0$
this gives
\begin{eqnarray}
&&\hspace{-10mm} V(x,\zeta,t) = \frac{1}{8\pi^2} \left (
 \theta[x(\zeta\!-\!x)]\!\!\int_0^{\frac{x}{\zeta}} \!\!  +
 \theta[(x\!-\!\zeta)(1\!-\!x)] \!\!\int_0^{\frac{1-x}{1-\zeta}}
 \right )\! d y D(x\!-y \zeta, y),\; (\zeta \ge 0),
\label{Fp}
\end{eqnarray}
with the first term having the support $x \in [0,\zeta]$, and the second 
$x \in [\zeta,1]$. For the case $\zeta < 0$ we obtain
\begin{eqnarray}
&& \hspace{-10mm} V(x,\zeta,t) =  \frac{1}{8\pi^2} \left ( 
\theta[x(\zeta-x)]\!\!\int_{\frac{x}{\zeta}}^{\frac{1-x}{1-\zeta}} \!\! +
 \theta[x(1-x)]\!\! \int_0^{\frac{1-x}{1-\zeta}} \right )
 d y  D(x-y \zeta,y),\; (\zeta < 0),
\label{Fm}
\end{eqnarray}
with the support $x \in [\zeta,0)$ for the first term and $x \in
[0,1]$ for the second term. Thus the support is correct.  As expected,
the function $V$ is continuous in the $x$ variable, with the
derivative $dV/dx$ discontinuous at the points $x=0,\zeta, 1$.

The contribution from the crossed diagram is formally obtained from
the direct diagram with the replacement and $p \to -p - q$. Replacing
correspondingly $x \to \zeta-x$ and performing the {\em M\"unchen}
transformation: $z\to z$, $y \to 1-y-z$, yields the result
${\bar V}(x,\zeta,t)=V(\zeta-x,\zeta,t)$.
The support of the crossed diagram reflects the support of the direct
diagram. For the case $\zeta \ge 0$ it is $x \in [-1+\zeta,\zeta]$,
while for $\zeta < 0$ we have $x \in [-1+\zeta,0]$.

The evaluation of the axial TDA proceeds analogously, with the trace yielding an
additional proportionality factor $(2 k + q ) \cdot \varepsilon$ as compared to the vector TDA
in Eq.~(\ref{Vgen}). The result is
\begin{eqnarray}
A(x,\zeta,t) = \frac{1}{8\pi^2} \int_0^1 d y \int_0^1 d z \delta \left
( x - z- y \zeta \right ) (2y+2z-1)D(z,y). \label{bubA}
\end{eqnarray}
In the present case
${\bar A}(x,\zeta,t)=-A(\zeta-x,\zeta,t)$,
with the minus coming from the trace factor.

Expressing the TDAs through the double distributions has the known
advantage of an automatic verification of the {\em polynomiality
conditions} which ultimately correspond to proper implementation of
the Lorentz invariance. Polynomiality is manifest from this form, as the
moments $\int dx V(x,\zeta,t) x^n$ involve integrals with the factor
$(z +y \zeta)^n$ and result in a polynomial in $\zeta$ of order $n$.
We end these general considerations by writing down the double
distributions in the so called symmetric variables, $\alpha$, and
$\beta$ (not to be confused with the previously introduced Feynman
parameters), defined as $y=(1+\alpha-\beta)/2$, $z=\beta$. For the
vector case we find
\begin{eqnarray}
&&\hspace{-7mm}D(\alpha,\beta)=\frac{\theta(1-|\alpha|-\beta) \theta(\beta) \omega^2}{\omega^2+ [1-(\alpha-\beta)^2]q^2/4 +\beta(1-\beta)p^2 + (1+\alpha-\beta) \beta p \cdot q}, \label{Vdoubleab}
\end{eqnarray}
while for the axial case we find $(\alpha+\beta) D(\alpha,\beta)$, in full 
agreement with Ref.~\cite{Tiburzi:2005nj}. The theta functions provide the standard integration limits
$\int_0^1 d\beta \int_{-1+|\beta|}^{1-|\beta|} d\alpha$. 

The general expressions presented above are formal, as quark models
require regularization.  The choice of a {\it finite} regularization
is far from trivial, as requirements of the Lorentz and gauge
invariance, as well as preservation of chiral Ward identities and in
particular anomalies (crucial in the present study) impose severe and
tight constraints. An elegant way of imposing a regularization which
obeys these requirements is achieved in the Spectral Quark Model (SQM)
\cite{RuizArriola:2003bs}. In this model, developed in the spirit of
the early work of Efimov and Ivanov \cite{efimov}, the quark mass
$\omega$ is treated as a spectral parameter of a generalized Lehmann
representation, which is integrated along a suitably chosen {\em
complex} contour $C$. The chirally symmetric effective action
constructed in Ref.~\cite{RuizArriola:2003bs} reads
\begin{eqnarray}
\Gamma[U,v,a] =-i N_c \int_C d \omega \rho(\omega) {\rm Tr} \log
\left ( i\slashchar{\partial} - \omega U^5  - \left(
\slashchar{v} + \slashchar{a} \gamma_5 \right ) \right ),
\label{eq:eff_ac} 
\end{eqnarray} 
where the trace for a bilocal (Dirac- and flavor-matrix valued)
operator $A(x,x')$ is given by $ {\rm Tr} A = \int d^4 x \,{\rm tr}
\langle A(x,x) \rangle$ with ${\rm tr}$ denoting the Dirac trace and
$\langle \, \rangle $ the flavor trace. The matrix $ U^5 = e^{ i
\gamma_5 \vec{\tau}\cdot \vec{\pi} /f} = \frac12 ( 1 + \gamma_5 ) U +
\frac12 ( 1 - \gamma_5 ) U^\dagger$, while $U = e^{ { i} \vec \pi
\cdot \vec \tau / f }$ is the flavor matrix representing the
pseudoscalar pions in the nonlinear representation. The symbols $v^\mu $ and
$a^\mu$ represent external vector and axial currents. In
Eq.~(\ref{eq:eff_ac}) the spectral density $\rho(\omega)$ acts as a
regulator. Actually, the expressions for one-quark-loop observables in
SQM are obtained from the preceding expressions by integrating over
$\omega$ with the spectral density $\rho(\omega )$. For a generic
observable $F$ we have
\begin{equation}
F_{\rm SQM}=\int_{C}d\omega \rho (\omega )F(\omega ),
\end{equation}
where $F(\omega )$ is the quark-model result with the quark mass set
to $\omega$.  In the meson-dominance version of SQM
\cite{RuizArriola:2003bs} we have
\begin{eqnarray}
\rho (\omega )&=& \rho _{V}(\omega )+\rho _{S}(\omega ),
\label{mesdom} \\ \rho _{V}(\omega )&=&\frac{1}{2\pi i}\frac{1}{\omega
} \frac{1}{(1-4\omega^{2}/M_{V}^{2})^{5/2}}, \; \rho _{S}(\omega )
=-\frac{1}{2\pi iN_{c}M_{S}^{4}}\frac{48\pi ^{2}\langle
\bar{q}q\rangle }{(1-4\omega ^{2}/M_{S}^{2})^{5/2}},\nonumber
\end{eqnarray}%
where $M_{S}=M_{V}=m_{\rho }$ is the $\rho$-meson mass
\cite{Megias:2004uj}.  The contour $C$ for the spectral integration
over $\omega $ and other details are given in
Ref~\cite{RuizArriola:2003bs}.  SQM generates the monopole
pion electromagnetic form factor \cite{RuizArriola:2003bs}.
Interestingly, the model has the feature of the analytic quark
confinement, {\em i.e.} the quark propagator has no poles, only cuts,
in the complex momentum plane. Moreover, the evaluation of low-energy
matrix elements in SQM is very simple and leads to numerous results
reported in Ref.~\cite{RuizArriola:2003bs}, in particular for the pion
light-cone wave function, the pion structure function, the generalized
parton distributions of the pion \cite{Broniowski:2003rp}, low energy
chiral Lagrangeans \cite{Megias:2004uj} and the photon, and
$\rho$-meson structure functions \cite{Dorokhov:2006qm}.

\begin{figure}[tb]
\begin{center}
\includegraphics[width=14.3cm]{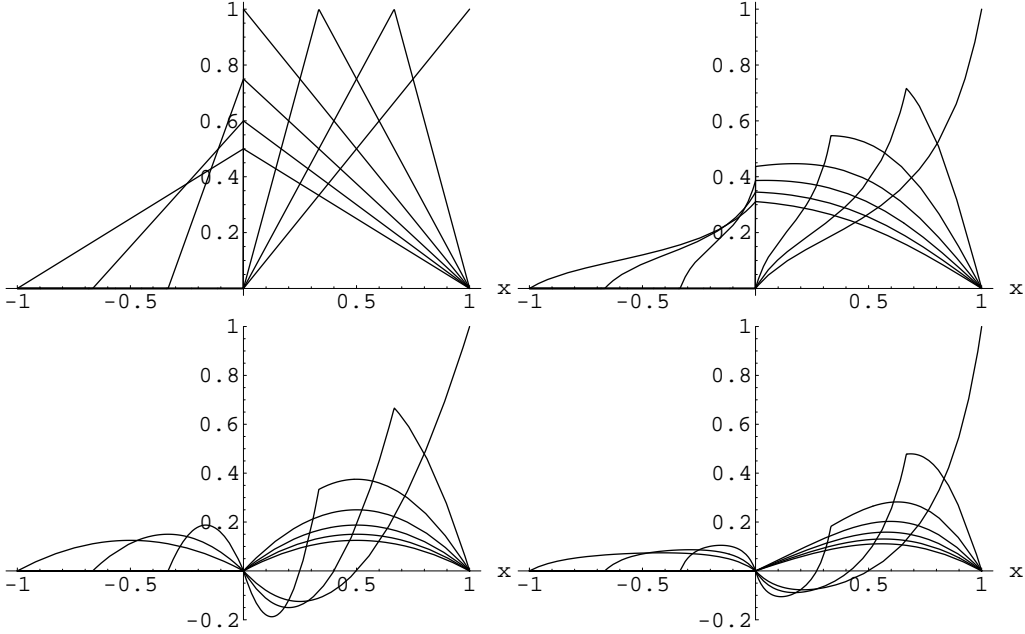} 
\end{center}
\vspace{-5mm}
\caption{Top: vector TDA $8\pi^2 V_{\rm SQM}(x,\zeta,t)$ for $t=0$
(left) and $t=-0.4$~GeV (right) plotted as functions of $x$ for
several values of $\zeta$: $-1$, $-2/3$, $-1/3$, $0$, $1/3$, $2/3$,
and $1$. The value of $\zeta$ can be inferred from the position of the
cusp for $\zeta>0$ or from the support for $\zeta<0$. Bottom: the same
for the axial TDA $8\pi^2 V_{\rm SQM}(x,\zeta,t)$. The vector curves
are normalized to $1/2$, and the axial curves to $1/6$.  The presented
calculation is made in the Spectral Quark Model with $M_V=0.77$~GeV in
chiral limit and for the real photon.
\label{fig:TDA}}
\end{figure}

Let us assume for simplicity from now on the chiral limit, $p^2=0$,
and the real photon, $(p+q)^2=0$.  The evaluation of the spectral
integral is straightforward using the results of Ref.~\cite{Megias:2004uj}. Then the double
distribution for the vector TDA in SQM assumes the simple form
\begin{eqnarray}
{D}_{\rm SQM}(z,y)=\int_C d\omega \rho(\omega) {D}(z,y) =
\frac{\theta(1-y-z)} {\left ( 1 - \frac{4 y(1-y-z)t}{M_V^2} \right
)^{5/2}}.
\end{eqnarray}
Remarkably, completely analytic expressions for the TDAs follow,
fulfilling all {\it a priori} Lorentz, chiral and gauge invariance
constraints. To our knowledge this is the first explicit calculation
of TDAs in a regularized chiral quark model, and we list the results
in some detail as their form may guide the used parameterizations of
TDAs. We only show the results for the case $\zeta \ge 0$, as for
$\zeta \le 0$ they have a similar character. We find for the vector
TDA
\begin{eqnarray}
V_{\rm SQM}(x,\zeta,t) &=& \frac{M_V^2}{48 \pi^2} \left (
\theta[x(\zeta-x)] \left (\chi_1+\frac{\chi_2}{2}\right ) +
\theta[(1-x)(x-\zeta)] \chi_2 \right ), \label{chi}
\end{eqnarray} 
where
\begin{eqnarray}
&&\hspace{-9mm} \chi_2=\frac{2 (x-1) \left[ 3 (\zeta-1) M_V^2+t
   (x-1)^2\right ]}{\left[ (\zeta-1) M_V^2+t (x-1)^2\right]^2}, \\ &&
   \hspace{-9mm} \chi_1= \frac{(x (\zeta -2)+\zeta ) \left(3 M_V^2
   (\zeta -1) \zeta ^2+t \left(\left(\zeta ^2+8 \zeta -8\right) x^2+2
   (4-5 \zeta ) \zeta x+\zeta ^2\right)\right)}{\left((\zeta -1)
   M_V^2+t (x-1)^2\right)^2 \left({\zeta^2+\frac{4 t x (x-\zeta
   )}{M_V^2}}\right)^{3/2}} \nonumber.
\label{chi1-chi2} 
\end{eqnarray}
Some special values are
\begin{eqnarray}
V_{\rm SQM}(0,\zeta,t)&=&V_{\rm SQM}(1,\zeta,t)=0, \;\;
V_{\rm SQM}(\zeta,\zeta,t)= \frac{M_V^2 \left( 3M_V^2+t (\zeta-1)\right)}{24 \pi^2 \left(M_V^2+t (\zeta-1)\right)^2},  \nonumber \\
V_{\rm SQM}(x,\zeta,0) &=& \frac{1}{8 \pi^2}\left ( \theta[x(\zeta-x)]
\frac{x}{\zeta} + \theta[(1-x)(x-\zeta)] \frac{1-x}{1-\zeta} \right ),
\nonumber \\
V_{\rm SQM}(x,0,t) &=& \frac{M_V^2 \left(3M_V^2-t (1-x)^2\right) (1-x)}{24 \pi^2 \left(M_V^2-t (1-x)^2\right)^2}. 
\end{eqnarray}
The functions $8\pi^2 V_{\rm SQM}(x,\zeta,t)$ are shown at the top of
Fig.~\ref{fig:TDA} for two values of $t$ and for several values of
$\zeta$, both positive and negative.
The integration over $x$ produces the $\zeta$-independent (as required
by polynomiality) form factors,
\begin{eqnarray}
&&\hspace{-10mm}\int dx V^{I=0}_{\rm SQM}(x,\zeta;t)\!=\!\int dx \left(V_{\rm SQM}\!+\!{\bar V}_{\rm SQM}\right)=\frac{M_V^2}{24 \pi^2}
\left( \frac{2}{M_V^2-t}-\frac{\log \left(1-\frac{t}{M_V^2}\right)}{t}\right).\nonumber \\
&&\hspace{-10mm}\int dx V^{I=1}_{\rm SQM}(x,\zeta;t)=\int dx N_c\left(V_{\rm SQM}-{\bar V}_{\rm SQM}\right)=0.  \label{ffF}
\end{eqnarray}
The isoscalar form factor is related to the pion-photon transition
form factor,
\begin{eqnarray}
F_{\pi \gamma \gamma^\ast}(t)=\frac{2}{f}\int dx V^{I=0}=
\frac{1}{4\pi^2 f}\left ( 1+ \frac{5t}{6M_V^2} + \frac{7t^2}{9M_V^7}
\dots\right ),
\end{eqnarray}
where the factor of 2 comes from the fact, that either of the photons
can be isoscalar. We read out the corresponding rms radius to be
$\langle r^2 \rangle^{1/2}_{\pi \gamma \gamma^\ast}=\sqrt{5}/M_V= 0.57
{\rm fm}$ for $M_V= 770 {\rm MeV}$. Equivalently, one may use the slope parameter 
$b_\pi=
\frac{d}{d t} F_{\pi^0 \gamma \gamma^\ast} (t)/F_{\pi^0 \gamma \gamma^\ast} (t)\Big|_{t=0}$.
SQM gives $b_\pi = 5/(6 M_V^2) = 1.4~{\rm GeV}^{-2}$, in 
very reasonable agreement with the experimental
values quoted by the PDG~\cite{Eidelman:2004wy}: $b_\pi = (1.79 \pm 0.14 \pm
14) {\rm GeV}^{-2}$ originally reported by the CELLO
collaboration~\cite{Behrend:1990sr}, 
$b_\pi = (1.4 \pm 1.3 \pm 2.6) {\rm GeV}^{-2}$ from
Ref.~\cite{Farzanpay:1992pz}, or $b_\pi = (1.4 \pm 0.8 \pm 1.4) {\rm
GeV}^{-2}$ given in \cite{MeijerDrees:1992qb}.

The vector form factor is also related to the $F_V$ form factor
appearing in the radiative pion decays, $\pi^\pm \to l^\pm \nu_l
\gamma$ (for a review see e.g. \cite{Bryman:1982et} and references
therein). With the assumption of CVC it is related to the the isoscalar
form factor in the following way:
\begin{eqnarray}
F_V(t)=\frac{\sqrt{2} m_\pi}{f} \int dx V^{I=0}=\frac{\sqrt{2} m_\pi}{8\pi^2 f}\left ( 1+ \frac{5t}{6M_V^2}+\dots \right ).
\label{fv}
\end{eqnarray}
The premultiplying factors follow the assumed conventions\footnote{We
follow~\cite{Eidelman:2004wy,Bryman:1982et}. The structure
dependent amplitude is given by $M_{\rm SD} = i e G \cos \theta_c \bar
u_\nu \gamma_\mu ( 1 - \gamma_5 ) v_e \epsilon^*_\nu M^{\mu \nu } /
\sqrt{2} m_\pi $, where the hadronic contribution is given by 
$ M^{\mu \nu} = F_V(t) \epsilon^{\mu \nu \alpha
\beta} p_\alpha q_\beta -i F_A(t) [ q^{\nu} 
(q^\mu + p^\mu) - g^{\mu \nu} q \cdot (q+p) ]$. }.  The value at $t=0$ (for $f=93$~MeV) is $F_V(0)\simeq
0.026$, as listed in \cite{Eidelman:2004wy}. The experimental data
fall one standard deviation below this CVC prediction, with $F_V(0)=
0.017\pm 0.008$ \cite{Eidelman:2004wy}.

For the axial TDA the corresponding expressions are (for $\zeta \ge 0$) 
\begin{eqnarray}
A_{\rm SQM}(x,\zeta,t) &=& \frac{M_V^2}{48 \pi^2} \left ( \theta[x(\zeta-x)]\chi_3 + \theta[(1-x)(x-\zeta)] x \chi_2 \right ), \label{chi3}
\end{eqnarray} 
where $\chi_2 $ is given by Eq.~(\ref{chi1-chi2}) and
\begin{eqnarray}
&&\hspace{-9mm} \chi_3= \frac{(\zeta -1)^2 M_V^4+t (x-1) (5 x-2)
   (\zeta -1) M_V^2+t^2 (x-1)^3 (2 x-1)}{t \left(t (x-1)^2+(\zeta -1)
   M_V^2\right)^2}+ \nonumber \\ && \frac{1}{t \left( M_V^2+\frac{4 t
   x (x-\zeta )}{\zeta ^2} \right )^{3/2} \left( t (x-1)^2+(\zeta -1)
   M_V^2 \right)^2} \\ && \times \left [ -(\zeta -1)^2 \zeta ^3
   M_V^4+t (\zeta -1) \zeta ^2 \left((\zeta -6) x^2+7 \zeta x-2 \zeta
   \right) M_V^2+ \right .  \nonumber \\ && \left . t^2 \left(16
   x^4-24 (x+1) \zeta x^3+6 (x (x+6)+1) \zeta ^2 x^2-(5 x (x
   (x+3)-1)+1) \zeta ^3\right) \right ]. \nonumber
\end{eqnarray}
The special values are
\begin{eqnarray}
&&\hspace{-9mm} A_{\rm SQM}(0,\zeta,t)=A_{\rm SQM}(1,\zeta,t)=0, \nonumber \\
&&\hspace{-9mm} A_{\rm SQM}(\zeta,\zeta,t)= \frac{\zeta M_V^2 \left( 3M_V^2+t (\zeta-1)\right)}{24 \pi^2 \left(M_V^2+t (\zeta-1)\right)^2}= 
\zeta V_{\rm SQM}(\zeta,\zeta,t), \nonumber \\ 
&&\hspace{-9mm}A_{\rm SQM}(x,\zeta,0) = \frac{1}{8 \pi^2}
\left ( \theta[x(\zeta\!-\!x)] \frac{x(\zeta x+x-\zeta)}{\zeta^2} + \theta[(1\!-\!x)(x\!-\!\zeta)]
\frac{x(1\!-\!x)}{(1\!-\!\zeta)} \right ), \nonumber \\
&&\hspace{-9mm}A_{\rm SQM}(x,0,t) = \frac{M_V^2 \left(3M_V^2-t
(1-x)^2\right) (1-x)x}{24 \pi^2 \left(M_V^2-t (1-x)^2\right)^2}=x V_{\rm SQM}(x,0,t) .
\end{eqnarray}
The form factors are
\begin{eqnarray}
&&\hspace{-10mm}\int dx A^{I=0}_{\rm SQM}(x,\zeta,t)=\int dx
\left(A_{\rm SQM}+{\bar A}_{\rm SQM}\right)=0, \nonumber \\
&&\hspace{-10mm}\int dx A^{I=1}_{\rm SQM}(x,\zeta,t)\!=\!\int dx N_c
\left(A_{\rm SQM}\!-\!{\bar A}_{\rm SQM}\right)= -\frac{M_V^2 N_c}{24
\pi^2}\frac{\log \left (1-\frac{t}{M_V^2}\right)}{t}.\label{ffFA}
\end{eqnarray}
The bottom part of Fig.~\ref{fig:TDA} displays the function $8\pi^2
A_{\rm SQM}(x,\zeta,t)$.

\begin{figure}[tb]
\begin{center}
\includegraphics[width=7.5cm]{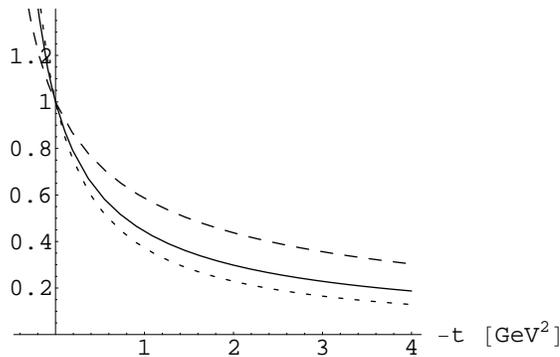} 
\end{center}
\vspace{-1cm}
\caption{Form factors $F_{\pi \to \gamma \gamma^\ast}(t)/F_{\pi \to
\gamma \gamma^\ast}(0)$ (solid line) and $F_A(t)/F_A(0)$ (dashed
line). The dotted line shows as reference the electromagnetic form
factor which in SQM is a monopole $F_\pi^{\rm em} (t) =
M_V^2/(M_V^2-t)$.
\label{fig:ff}}
\end{figure}

The axial TDA is related to the axial-vector form factor measured in
the radiative pion decays $\pi^\pm \to l^\pm \nu_l \gamma$ via
integration over the $x$ variable\footnote{This yields, according to
Eq.~(\ref{ATDA}), the quark contribution to the axial current from the
$\gamma^\mu \gamma_5$ vertex only, which by itself does not convey the
chiral Ward-Takahashi identity~\cite{RuizArriola:2003bs} at the quark
level. The additional pion pole contribution $-2 \omega q^\mu \gamma_5
/q^2 $ must be included in the vertex. Although this is essential to
preserve the identity $q^\mu \langle \gamma ( p' , \epsilon) |
J^{\mu,-}_A | \pi^+ (p) \rangle = i e ( \epsilon^* \cdot q ) \sqrt{2}
f_\pi $, and therefore for an unambiguous identification of $F_A(t)$
from the matrix element, for the tensor structure in Eq.~(\ref{ATDA})
such a contribution turns out to cancel. Hence in the evaluation of
the axial TDA we may retain only the pieces proportional to $p^\mu$ in
Eq.~(\ref{ATDA}), obtained with the $\gamma^\mu \gamma_5$ vertex.},
\begin{eqnarray} 
&&\hspace{-9mm}F_A(t)=\frac{\sqrt{2} m_\pi}{f} \int dx
A^{I=1}(x,\zeta,t)=\frac{\sqrt{2} N_c m_\pi}{24\pi^2 f }\left ( 1+
\frac{t}{2M_V^2}+ \frac{t^2}{3M_V^4}+\dots \right ),
\label{tranax}
\end{eqnarray}
where the premultiplying factors are the same as in Eq.~(\ref{fv}).
The value at $t=0$ is $F_A(0)=F_V(0) \simeq 0.026$, which is a factor
of 2 larger than the experimental number $F_A(0)= 0.0115\pm 0.0005$
\cite{Eidelman:2004wy}. The same conclusions were reached in
Ref.~\cite{Tiburzi:2005nj}. The predicted ratio
$F_A(0)/F_V(0)=1$ compares to the experimental number
of $F_A(0)/F_V(0)=0.7^{+0.6}_{-0.2}$ within one standard
deviation. The $t$-dependence of the form factors is presented in
Fig.~\ref{fig:ff}. We note long tails due to a $\log(-t) / t $
behavior at large $t$. The axial form factor (dashed line) lies above
the vector form factor (solid line), which in turn lies above the
monopole vector-meson dominance form factor, drawn as reference
(dotted line).

It is interesting to analyze our results in the light of Chiral
Perturbation Theory in the large-$N_c$ limit~\cite{Ecker:1989yg},
where $F_A(0) = 4 (L_{10}+ L_9) \sqrt{2} m_\pi / f_\pi = 0.012 \pm
0.008$ for $L_9 = (6.9 \pm 0.7)
\cdot 10^{-3} $ and $L_{10} = (-5.5 \pm 0.7) \cdot 10^{-3} $ (in fact
the values of the low-energy constants $L_9$ and $L_{10}$ are determined from 
the pion electromagnetic form factor and
the radiative pion decays).  In the large-$N_c$ limit one imposes QCD-motivated 
short-distance constraints regarding the asymptotic falloff
both for the difference of two-point vector minus axial correlators
(the Weinberg sum rules) as well as the axial and electromagnetic form
factors at large $t$ (see, {\em e.g.}, \cite{Pich:2002xy}). In the single
resonance approximation (SRA) one gets the same value for $F_V$ as
above and since $L_9^{\rm SRA} =-4L_{10}^{\rm SRA} /3 = f^2 / 2M_V^2 $
a ratio $F_A (0)/F_V (0) \sim 1/ 2 \sqrt{2} = 0.35 $ which produces a
significantly lower value for the axial form factor, $F_A^{\rm SRA} (0) \sim 0.01$. 
In SQM one has $L_{9}^{\rm SQM}=- 2 L_{10}^{\rm SQM} = N_c /(48 \pi^2)$~\cite{Megias:2004uj} 
and hence $ F_A^{\rm SQM} (0) = 0.026 $ in
agreement with Eq.~(\ref{tranax}).  The mismatch in the $L_{10}/L_9$
ratio in SQM and SRA stems from the absence of an explicit axial meson
contribution in SQM which induces the violation of the second Weinberg
sum rule pointed out previously~\cite{RuizArriola:2003bs} and
generating a $F_A(0)$ about twice the experimental number. This
feature is common to all known local quark models.
The SQM axial radius and the large-$N_c$ SRA result coincide, $\langle r^2 \rangle_A =
3 /M_V^2$, although at large $t$ SRA yields $F_A(t) \sim 1/t$ which is
slightly more convergent than our result (\ref{ffFA}), which contains an additional 
$\log(-t)$. 
It would be interesting to pursue the present calculation to the
nonlocal chiral quark models where both Weinberg sum rules are known
to hold \cite{Broniowski:1999dm}. Finally, the large $t$-behavior is
subjected to the QCD logarithmic radiative corrections.

Actually, the results obtained above hold at a {\em low energy scale} of the
quark model. In Ref.~\cite{Davidson:1994uv} an estimate of this scale
based on the momentum sum rule has been given. For the present model
or other local models, such as the Nambu--Jona-Lasinio model, the
scale turns out to be very low, around 320~MeV. An independent
estimate based on the pion light-cone distribution amplitude results
in a very similar estimate \cite{RuizArriola:2002bp}.  For that
reason, evolution is crucial for the case of DAs of PDFs, and
undoubtedly will also be important for the present case of TDAs as
well as the vector and axial form factors at large momenta. The
results obtained so far simply provide the initial conditions for the
QCD evolution, which at the leading order can be made with the usual
ERBL equations \cite{Efremov:1979qk,Lepage:1980fj}. A detailed study
of the evolution issues will be presented elsewhere.

In conclusion, we have presented an explicit quark-model study of the
vector and axial leading-twist pion-photon TDAs.  The results are
analytic, which allows us to gain insight into the possible forms of
non-perturbatively generated TDAs. Such predictions for TDAs or GPDs
are scarce, and frequently one only makes guesses subject to the
constraints coming from form factors, polynomiality, {\em etc.}  Our
method conforms to the Lorentz and gauge invariance, preserves the
chiral Ward-Takahashi identities as well as satisfies anomalies,
crucial in the study of the VAA processes. Since we proceed via the
double distributions, our TDAs automatically satisfy the polynomiality
constraints. The used technique of calculation, which takes advantage
of the $\alpha$-representation of propagators and thus is manifestly
covariant, makes a direct use of the Feynman diagrams and produces the
results in a few straightforward steps. No expansion in terms of
moments is necessary.  This technique is applicable also in other
calculations of this kind: PDFs, GPDs, as well as for other models,
including the non-local models, where the quark mass depends on the
virtuality. Our results correspond to a low-energy quark-model
scale. After suitable QCD evolution the obtained results may be used
in the studies of the virtual Compton scattering and other exclusive
processes involving pions, photons, and the weak gauge bosons.

We thank Brian Tiburzi for helpful e-mail exchanges and for 
pointing out a mistake in one of our formulas. 



\providecommand{\href}[2]{#2}\begingroup\raggedright\endgroup

\end{document}